%% file: aaskaii_template_V2.tex
\documentclass[a4paper,11pt]{article}
\usepackage{aaskaiid}
\usepackage{aas_macros}
\usepackage{tabularx}
\usepackage{makecell} 
\usepackage{booktabs}
\usepackage{tabularx}
\usepackage{hyperref}
\usepackage{lscape}
\usepackage{orcidlink}

\newcommand{\gcrt}{GCRT\,J1745\ensuremath{-}3009}

\title{Long-Period Transients as a new frontier in time-domain astronomy}
\ShortTitle{LPTs in the SKAO era}

\ShortName{Coordinators: Hao Qiu, Manisha Caleb} 

\author[1,2\dagger]{Manisha Caleb\orcidlink{0000-0002-4079-4648}}
\author[1,2]{Dougal Dobie\orcidlink{0000-0003-0699-7019}}
\author[3]{Natasha Hurley-Walker\orcidlink{0000-0002-5119-4808}}
\author[4]{Yogesh Maan\orcidlink{0000-0002-0862-6062}}
\author[5]{Yuan Mao\orcidlink{0000-0003-1874-0800}}
\author[6\dagger]{Hao Qiu\orcidlink{0000-0002-9586-7904}}
\author[7,8]{Nanda Rea\orcidlink{0000-0003-2177-6388}}
\author[1,9]{Kovi Rose\orcidlink{0000-0002-7329-3209}}
\author[1,2]{Iris de Ruiter\orcidlink{0000-0002-4752-5467}}
\author[10]{Ben Stappers\orcidlink{0000-0001-9242-7041}}
\author[3]{Ziteng Wang\orcidlink{0000-0002-2066-9823}}
\author[]{the Transients Science Working Group (In alphabetical order)}

\affiliation[1]{Sydney Institute for Astronomy, School of Physics, The University of Sydney, New South Wales 2006, Australia}

\affiliation[2]{Centre of Excellence for Gravitational Wave Discovery (OzGrav), Hawthorn, VIC 3122, Australia}
\affiliation[3]{International Centre for Radio Astronomy Research, Curtin University, Kent Street, Bentley WA, 6102, Australia}
\affiliation[4]{National Centre for Radio Astrophysics, Tata Institute of Fundamental Research, Post Bag 3, Ganeshkhind, Pune - 411007, India}
\affiliation[5]{National Space Science Center, Chinese Academy of Sciences, Beijing 100190, China}

\affiliation[6]{SKA Observatory, 26 Dick Perry Ave, Kensington WA 6151, Australia}

\affiliation[7]{Institute of Space Sciences (ICE-CSIC), Campus UAB, C/ de Can Magrans s/n, Cerdanyola del Vallès (Barcelona) 08193, Spain}
\affiliation[8] {Institut d’Estudis Espacials de Catalunya (IEEC), Esteve Terradas 1, RDIT Building, Of. 212 Mediterranean Technology Park
(PMT), 08860, Castelldefels, Spain}
\affiliation[9]{Australia Telescope National Facility, CSIRO, Space \& Astronomy, PO Box 76, Epping, NSW 1710, Australia}
\affiliation[10]{Jodrell Bank Centre for Astrophysics, University of Manchester}

\emailAdd{hao.qiu@skao.int}
\emailAdd{manisha.caleb@sydney.edu.au}

\abstract{Long-period radio transients (LPTs) are relatively new astrophysical objects occupying the observational gap between canonical pulsars and slowly varying radio variables. They emit coherent, highly polarised radio bursts with periods from minutes to hours, often exhibiting millisecond- to minute-scale substructure, short duty cycles, and broadband emission. Their radio luminosities typically exceed what rotational energy alone can power, necessitating alternative energy sources such as magnetic field decay, magnetospheric reconnection, or binary interactions.

As multiwavelength counterparts in X-ray, optical, and infrared bands provide key constraints on progenitors and emission mechanisms, observational evidence points to a diverse progenitor population including ultra-long period magnetars and magnetic white dwarf binaries. 
Fast imaging surveys with SKAO and its precursors are opening a new discovery space, enabling systematic detection, high-cadence monitoring, and detailed follow-up. Despite the challenges of high extinction, intermittent emission, and computational demands for discovery, the expanding LPT population provides a new laboratory for studying coherent radio emission in a range of compact-object systems, from pulsars to white dwarf binaries. This diversity allows us to test how the emission processes depend on magnetic field strength, rotation, and binary interaction.}

\begin{document}
\maketitle

\section{Introduction}

\begin{figure}[h]
    \centering
	\includegraphics[width=6.3in]{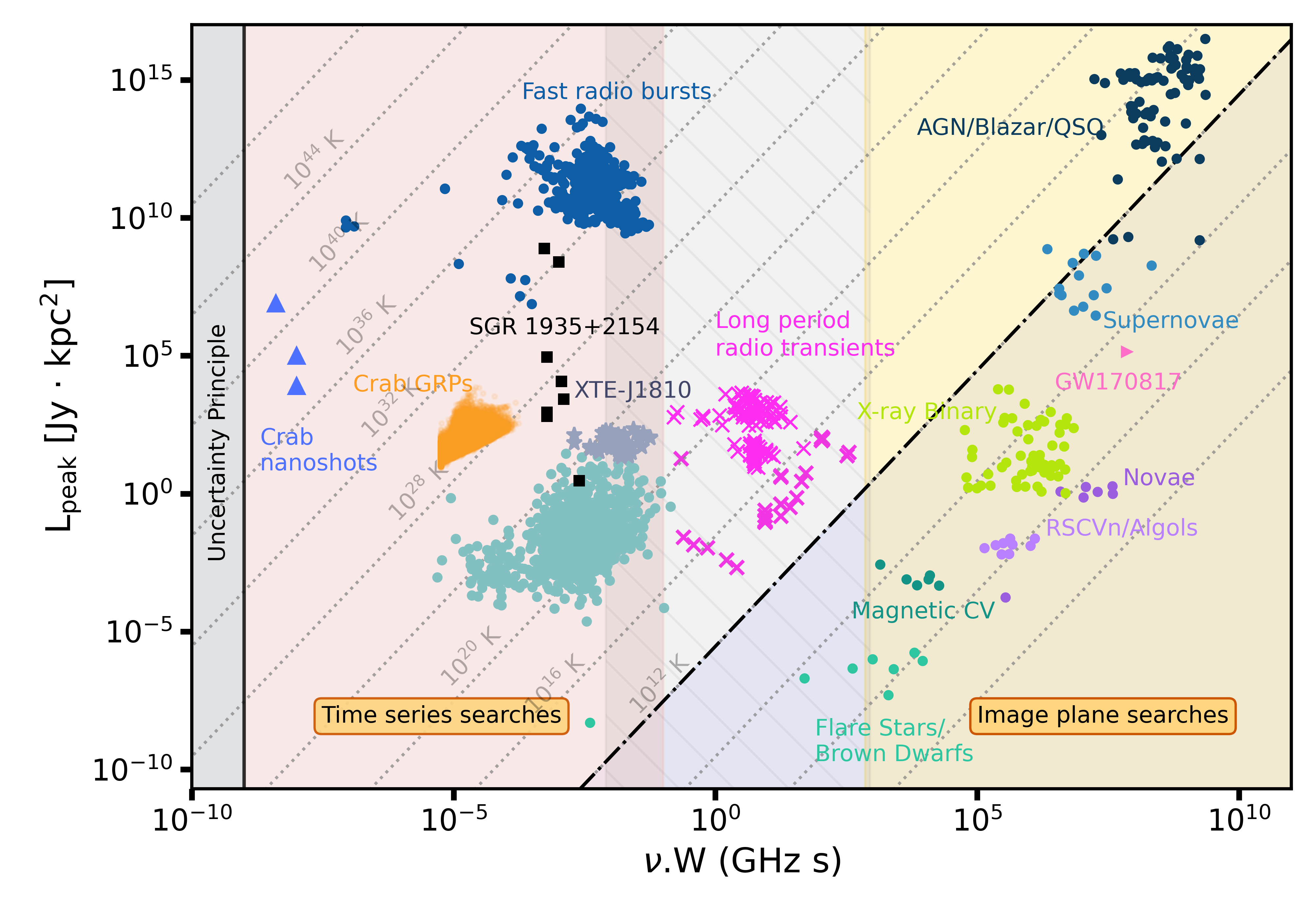}
    \caption{Radio transient phase space of various coherent and incoherent emitters. Diagonal lines represent constant brightness temperatures. The brightness temperature of 10$^{12}$ K separates coherent emitters from the incoherent ones, with the shaded region (lower right triangle) housing the incoherent emitters. The vertical shaded regions represent the timescales traditionally searched in time series surveys and image domain surveys. The hatched region represents the parameter space probed by fast imaging.}
    \label{fig:phasespace}
\end{figure}

Radio transients span a wide range of timescales and energies, from milliseconds bursts of coherent radio emission to explosions and afterglows that evolve over hours to years \citep{dawes-review} (also see Figure \ref{fig:phasespace}). Traditionally, time series searches and surveys \cite[e.g.][]{ransom+2002, cm2003, htru, superb} have been optimised to detect the millisecond-scale bursts and short-period signals, while image-plane searches and surveys \citep[e.g.][]{mooley+2016, lotss, vast} have focused on the much slower variables—such as supernovae and variable stars—evolving over timescales of hours to days. Between these regimes lies a largely unexplored window (Figure \ref{fig:phasespace}). Long-period radio transients (LPTs) occupy precisely this middle ground: their emission is too long in time to be efficiently detected by standard time series searches, yet too short and intermittent to appear significant in image stacks integrated over long durations. It is therefore unsurprising that such sources eluded discovery for decades. 

LPTs challenge our conventional understanding of compact object physics and population demographics. By studying them we gain insights into astrophysical processes such as binary interactions, accretion instabilities, and magnetospheric reconfigurations. This chapter reviews the emerging field of long-period transient science. We begin by defining the scope of what constitutes a `long-period' transient, then trace the historical development of this field. Subsequent sections will explore the current landscape of physical interpretations and its potential implications to binary/stellar evolution. In the final sections we discuss the future outlook of the field and the role of the SKAO telescopes.

\section{Observational Phenomenology of Long-Period Radio Transients}

LPTs are characterised by highly polarised coherent, relativistically beamed radio emission--often showing both strong linear and circular polarisation--with periods ranging from several minutes to many hours (see Tables~\ref{table:lpt-table} and \ref{tab:lpt-pol-table}). Their individual pulses typically last from a few seconds to several minutes often with microseconds to millisecond-duration substructure
(see references in Table~\ref{table:lpt-table}). In the radio band, their emission frequencies extend from $\sim100$ MHz to several GHz, with some objects detected across wide bandwidths, suggesting broadband coherent processes. LPTs were initially thought to be outlier neutron star pulsars, which was supported by the polarised coherent emission and lack of bright optical counterparts \citep{caleb2022discovery}. However, the radio luminosities inferred for these sources exceed their estimated spin-down luminosities (assuming magnetic dipole braking), implying that rotational energy alone cannot account for the observed emission \citep{hurley2023long}. 
This has prompted speculation that alternative energy reservoirs such as magnetic field decay, magnetospheric reconnection, or binary-driven accretion may play a role \citep[e.g.][]{cooper+2024, Qu+2025}. 

Some LPTs display stable emission over years to decades \citep{hurley2023long}, while others show irregular or intermittent behaviour lasting weeks to months \citep{hurley2022radio, caleb2024emission, Dobie2024}. A subset appear to reside in binary systems, confirmed through the detection of optical companions, where orbital modulation may influence the radio observed emission \citep{hurley20242, de2025sporadic}. For many others, the binary nature remains uncertain owing to the absence of any detectable optical counterparts. In a couple of cases, X-ray emission provides constraints on their energetics and environments \citep{wang2025detection, anumarlapudi2025askap}. 

Most LPTs have been identified since 2022 \citep{hurley2022radio}, but a few earlier unclassified radio transients may retrospectively fit the class. The best example is \gcrt\ \citep{hyman2005powerful}, discovered as five 1-Jy bursts during a 7-hour, 0.33-GHz VLA observation of the Galactic Centre in September 2002—hence the name Galactic Centre Radio Transient (GCRT). The bursts lasted $\sim$10 minutes and recurred every 77 minutes, with no high-energy counterpart detected \citep{hyman2002low}. Additional bursts were later reported by \cite{hyman2006new} and \cite{hyman2007faint}, but could not be phase-connected due to the long separation in time. 


LPTs have been extensively searched for emission beyond the radio band, with most sources yielding only upper limits in X-rays, optical, and infrared, likely due to distance or crowded fields (see Table~\ref{table:lpt-table}). Notably, ASKAP J1832$-$0911 \citep{wang2025detection} and ASKAP J1448$-$6856 \citep{anumarlapudi2025askap} show detectable X-ray emission, with ASKAP J1832$-$0911 exhibiting strong periodicity matching its 44-minute radio period. This suggests a connection between radio and X-ray emission with both likely originating from magnetically linked regions. Optical and infrared studies have confirmed counterparts for several LPTs, particularly GLEAM-X J0704$-$36 \citep{hurley20242} and ILT J1101$+$5521 \citep{de2025sporadic}, whose optical periods align with the radio periodicity and whose spectra show evidence consistent with white dwarf–low-mass star binaries \citep{rodriguez2025spectroscopic}.
Other candidates show tentative optical/infrared associations, sometimes with variability hinting at orbital effects or flaring. Overall, these observations highlight that LPTs can exhibit multiwavelength signatures, providing crucial constraints on their progenitors and emission mechanisms.

\begin{figure}[h]
    \centering
	\includegraphics[width=6.3in]{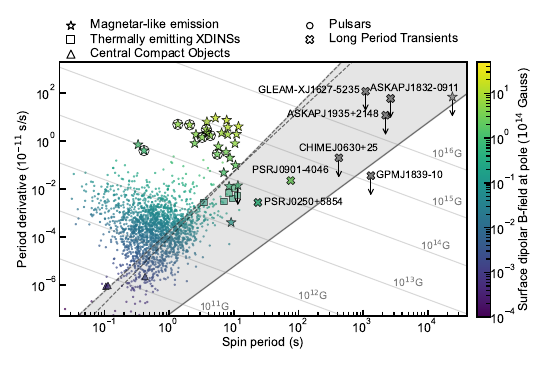}
    \caption{Period ($P$) versus period derivative ($\dot P$) diagram displaying LPTs alongside various pulsar populations. Downward arrows indicate upper limits on $\dot P$. Dashed and solid lines indicate theoretical death lines for a pure dipole and an extremely twisted multipole configuration, respectively. The colour scale represents the surface dipolar magnetic field strength at the pole.}
    \label{fig:lpt_ppdot}
\end{figure}

\include{lpt-table}

\include{lpt-pol-table}



\begin{figure}
    \centering
    \includegraphics{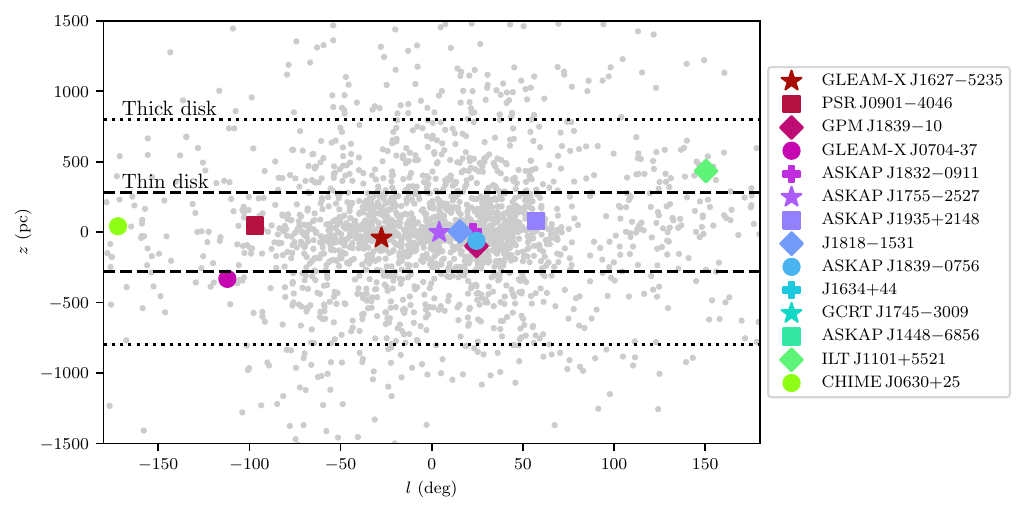}
    \caption{Spatial distribution of the 14 LPTs with distance estimates (colour) and known pulsars (grey; from \texttt{PSRcat} \citep{psrcat2005}) as a function of Galactocentric height and Galactic longitude. Dashed and dotted lines show estimates of the height of the thick and thin disks respectively. The current sample of LPTs is confined to the Galactic disk.}
    \label{fig:lpt-scaleheight}
\end{figure}

\section{Physical Origins and Models} \label{sec:physical_origins_and_models}



The coherent and high fractional linear polarised emission from LPTs suggests the presence of relativistic plasma and organised strong magnetic fields in the emission region.
Several progenitors have been proposed to explain the LPTs, including white-dwarf binary systems \citep{horvath2025unified, Qu+2025, yang2026} and long-period magnetars or neutron stars \citep{marsh+2016, pelisoli+2023, beniamini+2023, caleb2024emission}. 

Currently, there is no radio-only diagnostic that distinguishes between the progenitor scenarios. Multiwavelength observations along with theoretical modelling could help in pinning down the progenitors and understanding the nature of the emission. At present, the growing number of associations and modelling results suggests that white dwarf binary systems are a promising and increasingly favoured explanation, though the diversity of the LPT population leaves room for alternative progenitor channels.

\subsection{White Dwarf Binary Systems}
Detections of low-mass binaries with possible white dwarf companions in two LPTs, GLEAM-X\,J0704$-$36 and ILT\,J1101$+$5521 strongly suggest that progenitors of some of the LPTs belong to this class.
Systems such as AR Scorpii, and possibly these LPTs, represent detached pre-cataclysmic variable (CV) binaries where the white dwarf spin and orbital motion may already be synchronised. In some cases, as shown for GPM J1839$-$10 \citep{horvath2025unified}, the observed radio periodicity corresponds to the beat between the spin and orbital frequencies, with emission possibly triggered by magnetic interactions between the white dwarf and its companion’s wind.

The similarities in timescales between LPTs and CVs suggest that these populations could represent different evolutionary phases of magnetic white dwarf binaries. Many LPTs have orbital periods within the so-called CV ``period gap", where systems are detached and magnetic braking is inefficient. The relatively cool white dwarf temperatures in these LPTs indicate they have not undergone recent accretion, implying they are pre-CV remnants that have cooled for several billion years. Some short-period LPTs may only display the white dwarf spin or beat period rather than the full orbital modulation, depending on system geometry and observational sensitivity. It shows that LPTs may indicate an evolutionary stage between detached white dwarf–M dwarf binaries and fully accreting CVs.

\subsection{Magnetars}

An alternative explanation is that LPTs are powered by highly magnetised neutron stars, or magnetars, whose radio emission is known to be sporadic and polarised with quasi-periodic microstructure \citep{caleb2022discovery} and spectral volatility. The inter-pulses detected from ASKAP\,J183950.5$-$075635.0 and three different emission states observed from ASKAP\,J1935$+$2148 are highly reminiscent to similar properties seen in the emission from many of the Galactic neutron stars, and thus provide strong support to this particular progenitor class. Several other properties, e.g., orthogonal polarisation jumps, circular polarisation sign reversal, and correlated X-ray and radio emission provide additional support. The rotation energy loss alone is insufficient to explain the large energies of the brightest bursts observed from several LPTs, necessitating the radiation energy to be generated from another energy reservoir, such as the magnetic energy in magnetars.
Although typical magnetars rotate with periods of a few seconds, fallback accretion following a supernova could slow them to the very long periods observed in LPTs. A precedent exists in the 6.7-hour X-ray source 1E 161348$-$5055, thought to be a young magnetar braked by fallback material. This scenario could explain some radio/X-ray correlations seen in LPTs, but it faces difficulties — namely, the absence or presence of faint X-ray emission in most LPTs and the improbably high magnetar birth rate \citep[2.3 - 20 kyr$^{-1}$;][]{beniamini+2019}required to match the estimated population.

\subsection{LPTs in the context of other compact-object systems}

Beyond the white dwarf binary and magnetar scenarios discussed above, a number of additional interpretations have been proposed to explain the origin of LPTs. One connection has been drawn with long-period pulsars -- neutron stars with spin periods of tens of seconds \citep{tan2018lofar, wang2025discovery, caleb2022discovery} -- which occupy a transitional regime between canonical pulsars and the minutes- to hour-scale periodicities of LPTs. Their quasi-periodicity \citep{caleb2022discovery}, pulse-width scaling \citep{wang2025discovery}, and possible emission-state switching \citep{caleb2024emission} suggest possible phenomenological continuity, although their long spin periods challenge standard pulsar emission expectations.

A more tentative link has been suggested with repeating fast radio bursts (FRBs), motivated primarily by similarities in frequency–time structure observed in at least one LPT, GPM J1839$-$10 \citep{men2025highly}. At present, this resemblance is largely phenomenological, and establishing a physical connection will require systematic comparisons of energetics, polarisation properties, repetition behaviour, and burst morphology.

More speculative possibilities include isolated white dwarfs \citep[e.g.][]{LoebMaoz22}, binary neutron star systems undergoing propeller phases \citep[e.g.][]{mao+25}, and exotic compact objects such as strange dwarf pulsars \citep[e.g.][]{Zhou+2025}. See \cite{rea+2026} for a complete list of possible scenarios. While each scenario may account for specific aspects of the observed phenomenology, most face challenges in reproducing the full set of LPT properties, including luminosity, polarisation, and occurrence rates.

\section{Instrumentation and Survey Strategy for the SKA telescopes}

The recent surge in LPT discoveries has been driven by short-timescale (on seconds to minutes timescales) image-domain transient searches also known as `fast imaging'. 
Fast imaging with the SKAO presents an exciting opportunity for discovering more LPTs, as well as rapidly, automatically following them up (see Chapter 13). This approach has advanced in recent years thanks to improved processing capabilities and better understanding of SKA precursor and pathfinder data quality. Chapter 13 discusses these search methods in more detail. Short-timescale image-domain transient search pipelines have been developed for ASKAP \citep{wang2023radio}, LOFAR \citep{de2024transient}, MeerKAT \citep{smirnov2025mining}, and the MWA \citep{horvath2025long}. These pipelines have successfully identified LPTs as the search time scale matches the typical LPT burst duration. LPTs have also been identified in the CHIME single-pulse pulsar survey \citep{dong2024discovery, dong2025chime}. A complementary approach exploits the high circular polarization of LPT bursts, which makes them stand out in Stokes V images — where the sky is otherwise relatively empty making circularly polarized transients easier to identify \citep{bloot2025strongly, lee2025emission}.

\subsection{Cadence, Sensitivity Capabilities and the Role of AA4 in LPT searches}


The SKA telescopes will reach the original proposed design baseline capabilities (also known as SKA Phase 1) at the completion of Array Assembly 4 \citep[AA4;][]{sridhar2025ska_stage_delivery}. The SKA telescopes are expected to enter regular operations before that at with a preliminary Array Assembly known as  AA*  in 2030 for SKA-Low and 2032 for SKA-Mid.
SKA-Mid AA* will consist of 144 dishes and a maximum baseline of 108km, including 80 15m SKA antennas and 64 13.5m MeerKAT antennas; AA4 will extend to 159.6 km with 197 dishes. SKA-Low will include 307 stations (149 core stations) and reaching maximum 73.4 km baselines at AA*; AA4 will increase station number to 512 with the core consisting 224 core stations.



According to the delivery timeline, during AA* Cycle 0, the SKA telescopes will be able to search for new LPTs with the fast imaging pipeline and trigger the transient buffer to create high time resolution dynamic spectra for detailed analysis of new sources. With AA4 full capabilites, SKA-Low will be equipped with 400 core stations providing a 34\% increase of core sensitivity to AA*. SKA-Low will be able to conduct 0.8s snapshot cadence fast imaging searches at mJy level sensitivity. Shallow continuum surveys will provide 100 $\mu \mathrm{Jy}$ per 30 minute epoch images, which potentially enables the discovery of LPTs through traditional image-based radio transient surveys. While the extended 160 km baselines of SKA-Mid will allow sub-arcsec angular resolution observations for detailed follow-up studies.

With subarray and substation capabilities, the SKA will also be able to perform shallow large area sky surveys using specified sets of antennas/stations from the SKA telescopes. A sub-array consisting 16 SKA-Low stations will be able to search for LPTs above 7 mJy at 1-minute timescale resolution. SKA-Low will also be able to split into substations to increase the field of view.

The candidates generated from the fast imaging transient search may also be used to trigger other telescopes through distributed VOEvent alerts as part of the transient pipeline data product\citep{arumugam_2025_16950880} for multiwavelength follow-up, search for periodicity and monitor flaring activity.

In terms of discovery space, SKAO spans a wide range of frequencies, over which different LPTs have been seen to emit.  Some LPTs discovered so far have relatively steep spectra and are brighter at low frequencies; objects like this will be most easily discovered by SKA-Low. Scattering in the Galaxy will reduce the horizon out to which Low can be sensitive; since it is a function of $\nu^{-4}$, ionospheric refraction affects the data as $\nu^{-2}$, and sky noise is a function of $\nu^{-2.5}$, the top end of the SKA-Low band 350\,MHz is likely to be fruitful based on current detections from MWA, CHIME and LOFAR. 

Satellites effectively remove 240--270\,MHz from consideration. As long as station primary beams are well-behaved and the bandpass has no sharp drop-off at the top end, 270--320\,MHz is likely to be most effective for LPT searches (otherwise 200--250\,MHz). To reduce the impact of sporadic activity windows, field-of-view will be the most critical quantity to maximise. This can be achieved by using sub-stations.

\subsection{Survey Strategy}
Following the current specifications, the SKA-Low stations can be sub-stationed into 16 substations. The substations will enable increased field of view. If utilising the substations from each station to form 16 separate subarrays with different pointings, SKA-Low will be able tos cover $\sim1,250$\,sq.\,deg. at 300\,MHz, enough to completely cover the low-latitude Galactic Plane should it be overhead, or to perform an all-sky search, potentially with meridian drift scanning as per several searches with the MWA \citep[e.g.][]{2025PASA...42..129H}. 

To minimise computational cost and ionospheric effects (which will be more difficult to deal with for the less-sensitive, larger field-of-view sub-stationed array), it will likely be optimal to use a sub-array utilising stations within $<10$\,km of the core; the core itself is not necessary in its entirety due to the redundancy of the multiple short baselines and could be better utilised by other projects running simultaneously, such as pulsar timing or EoR searches. An array configuration like this should be $\sim10\times$ more sensitive than the MWA; depending on the steepness of the luminosity function and how many LPTs are transient (rather than persistent over decades)

On the other hand, some LPTs have a steep turnover and are invisible at low frequencies -- it is not currently known whether this is intrinsic or extrinsic to the sources \citep[see e.g.][]{wang2025detection,2025NatAs...9..393L}. There may be a population of faint LPTs; due to high scattering ($\propto\nu^{-4}$) in the Milky Way, these are more likely to be detected by SKA-Mid. In this case, field-of-view is difficult to increase as it necessitates sub-arraying, which then reduces the $u,v$-coverage, which could make it very difficult to search the complex Galactic plane (modulo, this would be trivial if SKAO enables a complete all-sky model early on in operations). Most likely, commensal fast-imaging with SKA-Mid would be the most effective search method, since it would be difficult to motivate any particular pointing, and for a faint Galactic population, almost any pointing is as good as any other (see Chapter 13 for more discussion).





\subsection{Real-Time Pipelines and Long-Term Monitoring}

Some LPTs are highly intermittent, exhibiting activity windows of days--weeks. Real-time discovery ensures radio follow-up to measure properties such as the period and dispersion measure can be carried out while the source is still active. In the SKA era this will be even more important --- raw visibilities will not be stored long-term, and hence any diagnostics that require them (higher time resolution imaging, dynamic spectra etc) must be created in real-time.

Current facilities take two approaches, through single pulse searches in dynamic spectra or image-based lightcurve subtraction. Single-pulse, real-time search backends designed for FRBs, such as CRAFT \citep{2010PASA...27..272M,2025PASA...42...36S} on ASKAP and MeerTRAP on MeerKAT \citep{tuse}, can discover new LPTs through their millisecond- to second-scale subpulse structures. However, the most successful LPT discovery technique in recent publications to-date is imaging model-subtracted visibilities at their native time resolution (typically 2--10\,s). This is mostly due to the high number of false positives in the subsecond phase space caused by impulsive RFI. Image/visibility plane model subtraction provide a more consistent averaged baseline and is still able to identify narrow sub-second bright pulses while opening the phase space to wider pulse-width bursts. While traditional real-time single pulse searches contain a relatively high number of false positives in the subsecond phase space from impulsive RFI. 

Fast-imaging processing has been recently incorporated into the real-time ASKAP CRAFT Coherent Upgrade \citep[CRACO][]{2025PASA...42....5W} workflow such as the new LOTRUN\footnote{\url{https://www.atnf.csiro.au/projects/science/wide-area-surveys/lotrun/}} survey, demonstrating that this approach may feasible in the SKA era with optimisation to scale. Similar fast-imaging approaches are adopted by MeerKAT \citep{smirnov2025mining}, the MWA \citep{horvath2025long} and LOFAR \citep{deruiter2024}. 
Fast imaging pipelines are generally commensal and utilise optimised resources parallel to the standard imaging pipelines to quickly produce candidates instead of traditional image-plane transient searches. 
Detailed descriptions of these fast-imaging pipelines can be found in the 'Commensal image plane transient searches with the SKA' chapter.



\section{Open Questions and Future Directions}

\subsection{Theoretical Gaps}

At the time of writing we do not conclusively know the progenitors of these sources, although there are several confirmed binary systems and several unclassified sources (Table~\ref{table:lpt-table}). The binarity of most systems is unclear, and is hindered by their low galactic latitude which impedes optical observations. In the binary progenitor scenario, we do not yet know the triggering mechanism for bursting activity, nor what differentiates them from other RS CVn or white dwarf binary systems. In the case of the isolated progenitor scenario it remains unclear whether these sources are simply the slow tail of the canonical pulsar/magnetar population or an exotic sub-class of neutron star. In short, the ``LPT'' moniker is currently entirely phenomological and our inability to link it to a physical model is a fundamental gap in our understanding.

However, this is a new but rapidly expanding class of object, and closing this gap simply requires a larger observational population. Detection of more low-extinction LPTs will allow us to determine the fraction of them that arise from binaries. A larger sample of intermediate-timescale sources with periods of tens of seconds (\citealp[e.g.][]{caleb2022discovery}) will shed light on the tail of the pulsar period distribution and the beginning of the LPT period distribution, and help determine the progenitors of the isolated LPTs. Long-term monitoring of specific sources will allow us to probe the evolution of their pulse profiles and in turn, their emission mechanisms.

\subsection{Observational Challenges}


The discovery of this source class primarily arose from the application of new observational techniques, so it is not surprising that there are still many observational hurdles ahead. Some are fundamental (e.g. most sources being highly extincted, long periods, high intermittency and large pulse widths hindering accurate timing) but most are surmountable.

The primary challenge for searching LPTs in the SKA era is the need for a mature fast imaging pipeline. LPTs have so far generally been found by reprocessing archival visibilities with custom fast imaging pipelines. This will not be practical in the SKA era due to the data rates at high time resolution. This calls for the development of a reliable detection pipeline that can produce lightcurves and full-Stokes dynamic spectra for candidates of interest with high accuracy; 
Such tools have already been developed for SKA precursor instruments, and have played a critical role in the discovery and study of LPTs.

Furthermore, the sub-second timescale behaviour of these sources is extremely vital -- quantifying pulse sub-structure and determining dispersion measures is generally not possible with standard visibilities and instead requires real-time formation of high time resolution visibilities \citep[e.g., CRACO;][]{2025PASA...42....5W} or simultaneous observations with pulsar backends (e.g. TUSE, PTUSE, MWA VCS) \citep{tuse, ptuse, mwa-vcs}. The Pulsar Subsystems (PSS and PST) will be able to provide such capabilities to record dynamic spectrum and baseband voltages for candidate detections, but may benefit from cross-pipeline triggering with the fast imaging pipeline.






\section{Outlook for the Next Decade}



With SKAO pathfinder facilities like ASKAP, LOFAR and MeerKAT now routinely performing fast-imaging in near real-time and already successfully yielding results, the field is moving from opportunistic discovery to systematic, survey-scale operation. 

Over the next decade we anticipate five key developments:

\begin{enumerate}
    \item \textbf{High-cadence surveys} — High cadence transient surveys utilising wide-field radio telescopes such as ASKAP’s commensal fast-imaging programme (including the upcoming LOTRUN\footnote{\url{https://www.atnf.csiro.au/projects/science/wide-area-surveys/lotrun/}} inner-Galactic-plane survey) and the MWA Galactic Plane Monitor project will deliver orders-of-magnitude larger sky-area coverage at sub-minute to minute image cadences, enabling thorough exploration of the LPT regime from minutes to hours. Looking further ahead, the SKAO telescopes will extend this capability to substantially higher sensitivities and finer angular resolution, opening up the faint, more distant LPT population and enabling Galaxy-wide and eventually extragalactic population studies.

    \item \textbf{Real-time pipelines and classification scale-up} - The DSA‑110 experience with its GPU‑accelerated ``cerberus" pipeline shows that short timescale image-plane single-pulse search in real time is feasible \citep{sherman+2025}. This underscores the imperative for large-scale, automated pipelines—not only to handle current surveys, but also to meet the even greater data-rate and candidate-classification demands that will be encountered by SKAO telescopes as they process vast data volumes, classify candidates, trigger follow-up, and filter false-positives.

    \item \textbf{Parameter-space extension and multi‐instrument synergy} - By extending the width sensitivity into the $\sim100-1000$~seconds regime (and beyond to multiple-hour timescales), the next decade will move into true ``long-period" territory for radio transients. SKAO telescope, with their unprecedented sensitivity and wide frequency coverage, will play a key role in probing fainter and more distant LPTs, while coordinated observations with existing radio facilities like ASKAP, MeerKAT, MWA, LOFAR and other facilities will maximise multi-instrument synergy.

    \item \textbf{Population studies and progenitor constraints} - The DSA-110 upper limits already rule out certain WD–M dwarf binary scenarios in the surveyed region at 95\% confidence \citep{sherman+2025}. Over the next decade the accumulation of non-detections, detections, and rate measurements (via ASKAP, DSA-110, MeerMAT, MWA, LOFAR and future instruments such as SKA‑Mid) will permit statistical population modelling of LPTs, identification of progenitor classes, and addressing of open questions.

    \item \textbf{Triggered follow-up and rapid multiwavelength response} - Real-time candidate generation will allow LPT detections to act as triggers for coordinated follow-up (optical, X-ray, gamma-ray). This capability is particularly important because some sources exhibit activity only during short, intermittent windows. The shift from archival searches to operational real-time discovery means we can target the source environment, capture transient multiwavelength counterparts, and place LPTs into a broader astrophysical context -- something that was rarely feasible in the ``legacy" era.

\end{enumerate}

In summary, the next decade represents a pivot in LPT research. With the SKA telescopes on the horizon, we expect the field of LPTs to transform from occasional discoveries into routine detections with real-time alerts and multiwavelength follow-up, thus revealing their origins and broader astrophysical roles.

\newpage 

\bibliographystyle{abbrvnat-maxbibnames4}
\bibliography{chapter10} 

\end{document}

%% file: lpt-table.tex

\begin{landscape}
\begin{table}[htb]
\renewcommand{\arraystretch}{1.2}
    \centering
    \scriptsize
    \begin{tabularx}{\linewidth}{p{2.5cm}lp{2.4cm}p{2cm}p{2.5cm}p{1.5cm}p{1.5cm}p{1.2cm}p{1.5cm}p{2.5cm}}
        \toprule
        Source Name & Period & P-dot & $L_{\rm radio}$ & $L_{\rm Xray}$ & OIR & Distance & Pulse Width  & Binary? & Remarks \& references\\
         & (mins) & (s/s) & (erg/s) & (erg/s) & (mag) & (kpc) & (s) & &  \\
        \midrule
GLEAM-X J162759.5$-$523504.3 & 18.18  & $<1.2\times 10^{-9}$     & $4 \times 10^{31}$ & $< 10^{32}$ & $> 23.7$ ($g$) & $1.3 \pm 0.5$ & 30 - 60 & No & [1,2] \\
PSR J0901$-$4046     & 1.26   & $2.25 \times 10^{-13}$   & --                 & $< 3.2 \times 10^{30}$ & $> 21$ & 0.3 - 0.5 & 0.3 & No & Pulsar/Magnetar [3] \\
GPM J1839$-$10       & 21.97  & $< 3.6 \times 10^{-13}$  & 10$^{28}$          & $< (0.1 - 1.5) \times 10^{32}$ & $19.7 \pm 0.2$ (?) & $5.7 \pm 2.9$ & 0.2 - 300 & Maybe & Active for 30 years. MD-WD binary [4,5,6]\\
GLEAM-X J0704‑37   & 174.94 & $< 3.9 \times 10^{-11}$  & $2 \times 10^{29}$ & $< 2.2 \times 10^{30}$ & --- & $1.5 \pm 0.5$ & 30 - 60 & yes & MD-WD binary? [7]\\
ASKAP/DART J1832$-$0911   & 44.2   & $< 9.0 \times 10^{-10}$  & $4 \times 10^{32}$ & $2.8 \times 10^{33}$ (peak) & $>19.98$ ($J$) & $4.5 \pm 1.2$ & 100 - 120 & Not clear & $L_{\rm X} \lesssim 6 \times 10^{31}$\,erg/s in quiescence [8]\\
ASKAP J175534.9$-$252749.1 & 69.8   &$(-10\pm92)\times10^{-12}$& --                 & $< (3-6) \times 10^{31}$ & $> 23$  & 4.7 & 50 - 150 & Not clear & Active for months [9,10]\\
ASKAP J1935+2148   & 53.76  &$(1.2\pm1.5)\times10^{-10}$& $(0.2-1.8)\times10^{30}$& $< 4 \times 10^{30}$ & $> 23.3$ ($g$) & 4.3 - 5.4 & 0.3 - 50 & Not clear &  [11]\\
ASKAP J183950.5$-$075635.0 & 387    & $1.6 \times 10^{-7}$    & $2 \times 10^{28}$ & $< 7.4 \times 10^{32}$ & $> 22.7$ ($g$) & 3.7 - 4.2 & 320 - 710 & Not clear & Inter-pulse [13] \\
CHIME/ILT J1634+44  & 14/70.1  &$-9 \times 10^{-12}$& $(0.2 - 4.0) \times 10^{29}$ & $< (0.5 - 1.3) \times 10^{32}$ & $> 25.4$ ($g$) & 1.0 - 4.3 & 1 - 5 & Not clear & Two periodicities [13,15]\\
GCRT J1745$-$3009    & 77     & --                      & -- &  &  &  & 600 &  &  [16]\\
ASKAP J1448$-$6856   & 93.9   & $< 2.2 \times 10^{-8}$  & $\sim 10^{28}\ast$ & ~10$^{29}\ast$ & 19.4 - 22.1 ($g$) & --- & 800 - 1800 & Not clear & $\ast$assume 1\,kpc distance [17]\\
ILT J1101+5521     & 125.5  & $< 1.7 \times 10^{-11}$ & $3.7 \times 10^{27}$ & $< 1.6 \times 10^{30}$ & 20.9 (r) & 0.5 & 30 - 90 & Yes & MD-WD binary [18]\\
CHIME J0630+25     & 7.02   & $5.2 \times 10^{-12}$   & -- & $< (4-15) \times 10^{29}$ & -- & 0.17 & 0.1 - 1 & Not clear &  [19]\\
        \bottomrule
    \end{tabularx}
    \caption{Properties of the known LPTs [references --- 1: \citet{hurley2022radio}; 2: \citet{Rea2022gleam}; 3: \citet{caleb2022discovery}; 4: \citet{hurley2023long}; 5: \citet{men2025highly}; 6: \citet{horvath2025unified}; 7: \citet{hurley20242}; 8: \citet{wang2025detection}; 9: \citet{Dobie2024}; 10: \citet{mcsweeney2025new}; 11: \citet{caleb2024emission}; 12: \citet{annathomas2024}; 13: \citet{lee2025emission}; 14: \citet{bloot2025strongly}; 15: \citet{dong2025chime}; 16: \citet{hyman2005powerful}; 17: \citet{anumarlapudi2025askap}; 18: \citet{de2025sporadic}; 19: \citet{dong2024discovery}.]. A catalogue of LPTs can be found at \href{https://lpt.mwa-image-plane.cloud.edu.au/published/tables/1}{https://lpt.mwa-image-plane.cloud.edu.au/published/tables/1.}}
    \label{table:lpt-table}
\end{table}
\end{landscape}


%% file: lpt-pol-table.tex
\begin{table}[htb]
\renewcommand{\arraystretch}{1.2}
    \centering
    \scriptsize
    \begin{tabularx}{\linewidth}{p{2.5cm}p{1.8cm}p{1.8cm}p{1.8cm}p{4cm}p{2cm}}
        \toprule
        Source Name & Linear \newline Polarisation & Circular \newline Polarisation & Rotation \newline Measure & Polarisation \newline Positional Angle & Reference \\
         & & & (rad\,m$^{-2}$) & & \\
        \midrule
        GLEAM-X J162759.5$-$523504.3 & $\sim$90\% & No & $-61\pm1$ & No change as a function of either the pulse phase or the observation time & [1] \\
        PSR J0901$-$4046 & $\sim$10\% & $\sim$20\% & $-64\pm2$ & S-shape & [3] \\
        GPM J1839$-$10 & $\sim$10--100\% & $\sim$10--100\% & $-531.83\pm0.14$ & flat within pulses, orthogonal polarisation jump & [4,5] \\
        GLEAM-X J0704$‑$37 & $\sim$20--50\% & $\sim$10--30\% & $-7\pm1$ & --- & [6] \\
        ASKAP/DART J1832$-$0911 & $\sim$80\% & $\sim$30\% & $+90.5\pm0.1$ & relatively flat within pulses, changes at the end of some pulses & [7] \\
        ASKAP J175534.9$-$252749.1 & $\sim$60\% & $\sim$40\% & $+961\pm45$ & S-shape & [8] \\
        ASKAP J1935+2148 & $\sim$40--90\% & <3\%--$\sim$70\% & $+159.3\pm0.3$ & flat across the pulse profile during the active state & [10] \\
        ASKAP J183950.5$-$075635.0 & 60--90\% (Main) \newline $\sim$90\% (Inter) & 30--60\% (Main) \newline $\lesssim$10\% (Inter) & +214 -- 219 & main pulses have a positive slope and interpulses are U-shaped (similar to PSR~B1702$-$19) & [12] \\
        CHIME/ILT J1634+44 & 0--100\% & 0--100\% & $-12$ -- $+11$ & --- & [13, 14] \\
        GCRT~J1745-3009 & --- & 0--100\% & --- & --- & [15] \\
        ASKAP~J1448-6856 & 35--100\% & 8\% & --- & --- & [16] \\
        ILT~J1101+5521 & 13--51\% & No & $4.72\pm0.04$ & large polarization position angle swing in one burst & [17] \\
        CHIME~J0630+25 & ? & ? & $-347.8\pm0.6$ & --- & [18] \\ 
        \bottomrule
    \end{tabularx}
    \caption{Polarization Properties of the Known LPTs}
    \label{tab:lpt-pol-table}
\end{table}